\documentclass{dynafront2025}

\usepackage{caption}
\usepackage{graphicx}
\usepackage{nicefrac}

\newcommand{\agent}{a}
\newcommand{\val}{v}

\newcommand{\probascenario}{\mu}
\newcommand{\DISCRETEBIDS}{\mathbb{B}}

\newcommand{\strategy}{{\gamma}}
\newcommand{\bid}{{b}}
\newcommand{\scenario}{S}
\newcommand{\agentprofit}{{\pi_\agent}}

\newcommand{\AGENT}{\mathbb{A}}

\DeclareMathOperator*{\argmax}{arg\,max}

\title[Fictitious Play in first-price auctions with correlated values]{An empirical study of Fictitious Play for estimating Nash equilibria in first-price auctions with correlated values}

\optauthor{\Name{Benjamin Heymann} \Email{b.heymann@criteo.com}\\
\addr Criteo AI Lab, Fairplay joint team
Paris, France
\AND
\Name{Panayotis Mertikopoulos} \Email{panayotis.mertikopoulos@imag.fr}\\
\addr CNRS, Laboratoire d’Informatique de Grenoble (LIG)
}

\begin{document}

\maketitle

\begin{abstract}This study concerns the computation of the \textit{Nash equilibria} of \textit{first-price auctions} with \textit{correlated values}.
Although some equilibrium computation methods exist for auctions with \textit{independent} values, the correlation of bidders' values introduces significant complications that render the existing methods unsatisfactory.
Our empirical contribution is a step towards filling this gap.
We report surprisingly good numerical convergence of \textit{Fictitious Play} toward an  $\epsilon$-equilibrium for an extensive set of instances. By doing so, we extend the insights of~\cite{rabinovich2013computing} to the correlated setting. These preliminary results call for further investigations into the properties of fictitious play algorithms on first-price auctions. 
\end{abstract}

\section{Introduction}

In single-item sealed bid first-price auctions (sometimes referred to as \textit{pay-as-bid}), several potential buyers bid simultaneously for an item.
The highest bidder then  obtains the item and pays the corresponding bid to the seller.
A common approach to auctions is to frame them as games. 
More precisely, because buyers’ values are not necessarily known to other participants, it is usually assumed that each of these values is a random variable sampled from a probability distribution.
In this narrative, each buyer is assumed to have formed their individual value \textendash\ a \emph{private} signal \textendash\ before submitting their bid, so that the resulting strategic interactions can be recast in the form of a \emph{Bayesian game}. A sizable part of the auction-theoretic literature makes an additional structural assumption that agents' values are independently distributed (an assumption we do not make).

Given the probability distributions of  buyers' values, an important research question is to determine the outcome of the auction.
While several solution concepts exist in the game-theoretic literature, the notion of (Bayes-)Nash equilibrium~\footnote{since the context is clear, we will use the term Nash equilibrium, or just equilibrium in this article} is of primary theoretical and practical importance for the study of  auctions, and    equilibrium numerical estimation has been a vigorously researched question for quite some time, with several breakthroughs along the way.
However, a major challenge that arises in practice is that
(\textit{a}) the above methods invariably rely on a first-order characterization of the solution
and
(\textit{b}) they require the bidders' values to be independently distributed, which is a very stringent limitation for real-life applications of auction theory (\textit{e.g.}, in online ad auctions).

In parallel, Fictitious Play (FP) was first discussed in~\cite{Bro51}. 
It has attracted considerable attention because (a) it is a simple procedure that  has been used to justify the Nash equilibrium play~\cite{brandtRateConvergenceFictitious}, and (b) despite being slow, it has been used in AI to approximate Nash equilibrium~\cite{ganzfried2008computing,rabinovich2013computing}.

\subsection*{Contributions}

In this work, we broaden the empirical results from~\cite{rabinovich2013computing} to accommodate auctions with correlated values, employing several technical adaptations (see Section~\ref{sec:xp}).
In all the examples we tested, the method approximated the  equilibrium surprisingly well.
Notably, the FP could be the first approach in the literature that remains agnostic to any correlations or dependencies between bidders' value distributions.

\section{Numerical solutions for auctions}
The literature on auctions is immense~\cite{krishna2009auction,Roughgarden_2016} and is impossible to survey in a short paper.
Therefore, we discuss below only the most relevant works we are aware of concerning the numerical computation of Nash equilibria in first-price auctions.

In his seminal paper~\cite{vickrey1961counterspeculation}, Vickrey derived a characterization of  equilibria when bidders' values are independent and otherwise identical.
Later, Plum~\cite{plum1992characterization} showed how to compute the  equilibrium of $2$-bidder auctions for some special cases when the price is a combination of the first and second prices.
The first general numerical method for computing the  equilibrium of a first-price auction with independent values appears in~\cite{marshall1994numerical}.
Theoretical analyses of the equilibrium structure are provided in particular in~\cite{10.2307/2648842,maskin2003uniqueness,reny2004existence}.
To study bidding rings, Bajari introduced several heuristics to compute the  equilibrium of first-price auctions~\cite{bajari2001comparing}.
Several other computation methods for first-price auctions with independent values have been proposed since then~\cite{Gayle2008,fibich2011numerical,kaplan2012asymmetric,Fibich2012,hubbard2014numerical}, and more recently,~\cite{10.5555/3398761.3398929}, which addresses the cases of discrete value distributions.
These methods rely on a first-order characterization of the players best replies to produce a system of ordinary differential equations, which are then solved using various methods.
One of the main difficulties lies in the numerical instability of the solutions. 

Research on first-price auctions received renewed impetus in 2019, when Google switched its display advertising marketplace to first-price auctions ~\cite{heymann2020bid,paes2020competitive}.
This led to a surge of interest in new topics such as 
computational complexity~\cite{filos2021complexity},
numerical approximation~\cite{rasooly2021importance}
or, more recently, neural networks  to compute auction equilibrium \cite{bichler2021learning}.
We also mention that there is an active track of research that, in the wake of~\cite{myerson1981optimal}, aims to maximize the seller's revenue~\cite{hartline2009simple,dhangwatnotai2015revenue,cole2014sample}.

\section{Fictitious play}

The work that is closest in spirit to our own is mostly empirical work \cite{rabinovich2013computing}, where the authors rely on FP running on a discrete set of possible actions.
That said, the setting of \cite{rabinovich2013computing} still differs  from our own in that bidders are assumed therein to be symmetric in most of the paper, and their values are further assumed to be continuously and \textit{independently} distributed.
In contrast, we consider both correlations between bidders (a crucial extension to capture real-life market behaviors) and atoms in the  value distributions.

The original fictitious play process was pioneered concurrently by Brown \cite{Bro51} and Robinson \cite{Rob51}.
This is one of the most widely studied procedures for learning in games, and it involves each player playing a best response to his/her beliefs about his/her opponents, given here by the  empirical frequency of past play.
Robinson first established the convergence of these beliefs to  equilibrium in two-player zero-sum games \cite{Rob51}.
Subsequently, the method has been shown to converge in
$2\times2$ games \citep{Miy61},
general $N$-player potential games \citep{MS96-jet},
symmetric games with an interior evolutionarily stable strategy \citep{Hof95b},
and certain classes of supermodular games \citep{MR90,MR91,Kri92,Hah08}.
Variants of FP involving a certain degree of explicit exploration/randomness have also been considered in the literature:
The most widely studied of these processes is that of \textit{stochastic} -- or \textit{perturbed} -- FP, which was introduced by Fudenberg \& Kreps \citep{FK93} and shown by Hofbauer \& Sandholm \citep{HS09} to converge to an approximate  equilibrium -- a quantal response equilibrium to be exact -- in the same classes of games as FP.\footnote{Stochastic FP is related -- \textit{but not in any way equivalent} -- to the class of no-regret learning policies known as "follow the regularized leader" \citep{SS11};
for a detailed discussion, we refer the reader to \cite{MS16,HLMS21a}, and references therein.}

At the same time, the literature on the convergence of (stochastic) FP should not be interpreted as suggesting that these processes converge to equilibrium in \textit{all} games.
Notable examples include Jordan's three-player matching pennies variant \citep{Jor93}, as well as the counterexamples by Shapley \citep{Sha64} and Gaunersdorfer and Hofbauer \citep{GH95}.
Similarly, we should stress that we do not make any claims of global convergence to  equilibrium in \textit{all} first-price auctions.
However, the series of numerical examples presented is sufficiently wide in scope and breadth to provide reasonable optimism for the practical use of FP, and call for further theoretical investigations.

\section{Auction model}

An auction is a Bayesian game where the players are the auction participants. 
    Each player places a bid after seeing their random and private value. Then, the auction is cleared according to the first-price auction rules.\footnote{In case of equality of the highest bids, a tie-breaking rule is needed.}.
    When the values and the bids belong to a finite set,  it is  practical to take a perspective based on agents, by associating each pair  (player, value)  to  an agent. Hence, we can  regard  the auction  as a tuple $(\AGENT,\mathcal{P},\val,\probascenario,\DISCRETEBIDS)$ such that 
$\AGENT$ is a finite set of agents --one per   (player, value) pair--, $\mathcal{P}$ maps back each agent from $\AGENT$ to the   player it represents, 
   $\val$  maps each agent from $\AGENT$ to a value,
  $\probascenario$ is  a probability distribution on 
$2^\AGENT$ that encodes the common knowledge Bayesian prior, and $\DISCRETEBIDS$ is a grid of admissible bids. 
A policy $\gamma_a$ for an agent $a$ is then a probability distribution on the set of bids $\DISCRETEBIDS$.

Let $S_{\agent}=\{S\in 2^{\AGENT}\mid\agent\in S\}$ be the set of all possible combinations of agents that contains agent $a$.
For a first-price auction,
the agent's (expected) payoff writes
\begin{align}
\pi_\agent(\strategy_\agent,\strategy_{-\agent})=\sum_{S\in S_{\agent}}\sum_{b\in \mathbb{B}^\AGENT} \hat{\pi}_\agent(b_a,b_{-a})
\mu(S)\strategy_{-\agent}(\bid_{-\agent})\strategy_{\agent}(\bid_{\agent}),
\end{align}
where $\hat{\pi}_{\agent}(b_a,b_{-a})$ is the payoff of  agent $a$ when their bid is $b_a$ and the other agents bids are  concatenated in the vector $b_{-a}$. In the absence of ties, this is 
$\hat{\pi}_{\agent}(b_a,b_{-a})=    (v_\agent-\bid_\agent)\prod_{\agent'\in\scenario\setminus \agent}\{\bid_\agent>\bid_{\agent'}\}$.
A tie-breaking rule needs to be specified  for the case of multiple highest bidders. It is typically the lexicographic order or a random allocation  among the highest bidders.
It seems empirically that giving a payoff of $0$ to everybody in case of ties improves the convergence of FP. While this is a choice that is hard to justify for an application, the impact on the game $\epsilon$-equilibrium\footnote{An $\epsilon$-equilibrium in this setting is a strategy profile
$\strategy^\star$ such that for all $\agent\in\AGENT$ and all alternative strategy profile $\strategy  $, $\agentprofit(\strategy^\star_\agent,\strategy_{-\agent}^\star)\geq
       \agentprofit(\strategy_\agent,\strategy_{-\agent}^\star)-\epsilon$.
 } is arguably small in many cases, so that this tie-breaking rule can be used to approximate the real game with FP.

\section{Experiments}
\label{sec:xp}

FP can be defined as follows for any stage $k>0$ 
\begin{align}
     &\bid_\agent^{(k)} \in \argmax_{\bid \in \DISCRETEBIDS}  \agentprofit(\delta_\bid,\strategy_{-\agent}^{(k)}), \quad \forall a\in \mathbb{A},\\
     &      \strategy_{\agent}^{(k+1)} = (1-\eta_k)\strategy_{\agent}^{(k)}+\eta_k\delta_{ \bid_\agent^{(k)}}, \forall a\in \mathbb{A},
\end{align}
where $\eta_k>0$ is a sequence of learning rates, and   $(\gamma^0_{\agent})_{\agent\in\mathbb{A}}$  an initial strategy profile, and $\delta_b$ is a mass of $1$ at $b$. 
We take $\eta_k=\nicefrac{1}{k+1}$ in most of the experiments. 
The algorithm and code for the experiments were implemented in Julia~\cite{bezanson2017julia} and can be found on this Github repository~\url{https://github.com/BenHey/FP4FPA}. 
Other experiments are  reported in a earlier version of the paper~\cite{heymann2021heuristic}. 

\begin{figure}
    \centering
    \includegraphics[width=1.\linewidth]{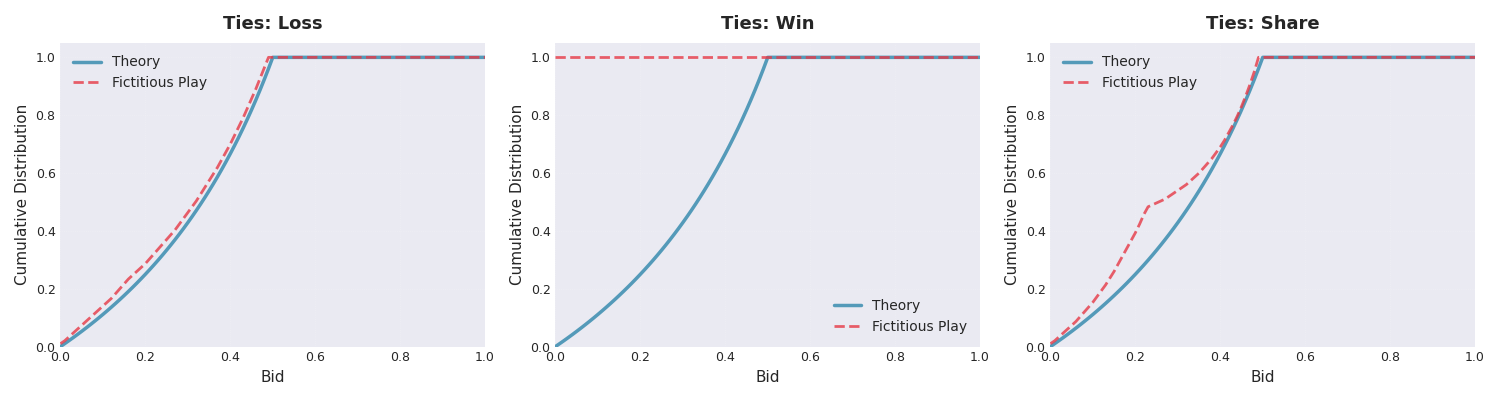}
    \caption{We consider a two-player setting where values are sampled independently and uniformly from $\{0,1\}$. The blue curve corresponds to the theoretical solution when bids can be in $[0,1]$ (continuous case). The tie-breaking rule we propose --- no player receives the item when bids are equal --- results in a reasonable approximation of the equilibrium (left). In the middle, we tested giving an item to each highest bidder, and on the right, to split the item in half.  }
    \label{fig:ties}
\end{figure}

 We recommend a tie-breaking rule, in which no player receives the item when bids are equal. Figure~\ref{fig:ties} provides an ablation study illustrating this approach. We consider a two-player setting where values are sampled independently and uniformly from $\{0,1\}$. When bids are constrained to $[0,1]$, the equilibrium bidding policy can be derived analytically: players with  value zero should bid $0$, while players with value $1$ should randomize their bids according to the cumulative distribution function $F(b) = \min(1/(1 - b) - 1, 1)$.
   We compare three tie-breaking rules. Only the one we propose, on the left, results  in a convergence close to the continuous, theoretical  equilibrium.

\begin{figure}
    \centering
    \includegraphics[width=0.5\linewidth]{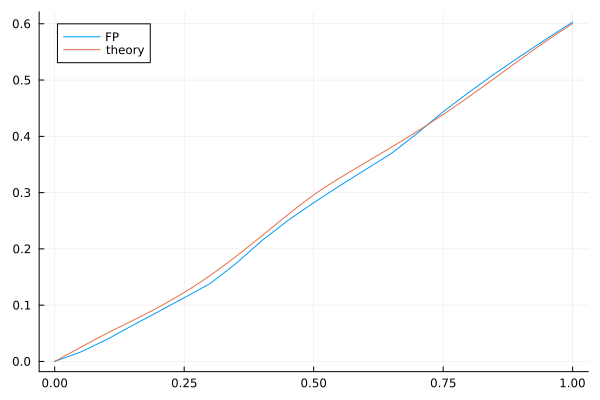}
   \caption{We consider an auction with two bidders, whose correlated values are distributed in $[0,1]$. 
        The generative model first tosses a coin $c$ valued in $\{L,H\}$. 
        Then, if $c=L$, the values are uniformly and independently sampled in $[0,2/3]$. Otherwise, if $c=H$, the values are sampled uniformly and independently in $[1/3,1]$.
        We can approximate the theoretical solution using numerical integration~\cite{milgrom1982theory}. We obtain $\epsilon=1.2\cdot 10^{-5}$}
    \label{fig:1}
\end{figure}

\begin{figure}
    \centering
    \includegraphics[width=0.5\linewidth]{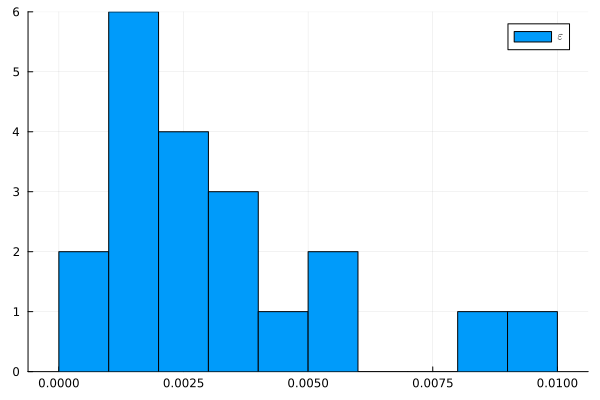}
   \caption{
     Distribution of $\epsilon$ for the 20 experiments on random environments.}    \label{fig:2}
\end{figure}

\begin{figure}[htbp]
    \centering
\includegraphics[width=0.3\textwidth]{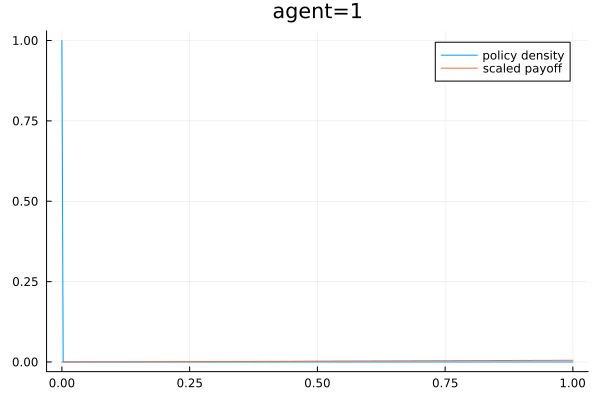}\includegraphics[width=0.3\textwidth]{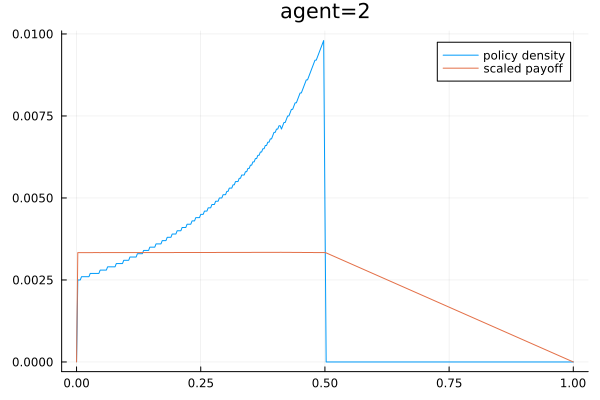}
\includegraphics[width=0.3\textwidth]{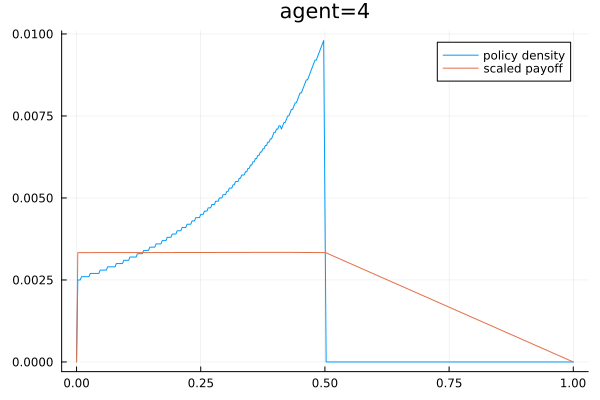}\\
\includegraphics[width=0.3\textwidth]{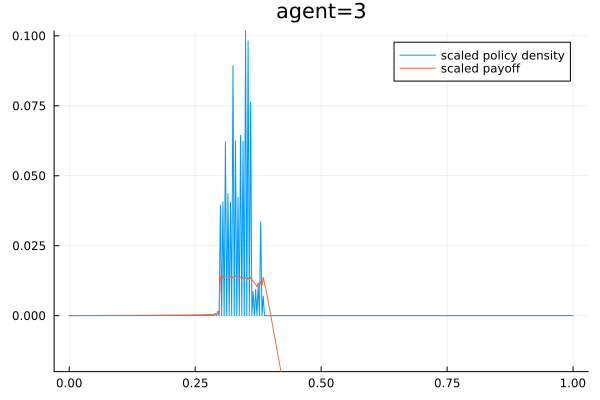}\includegraphics[width=0.3\textwidth]{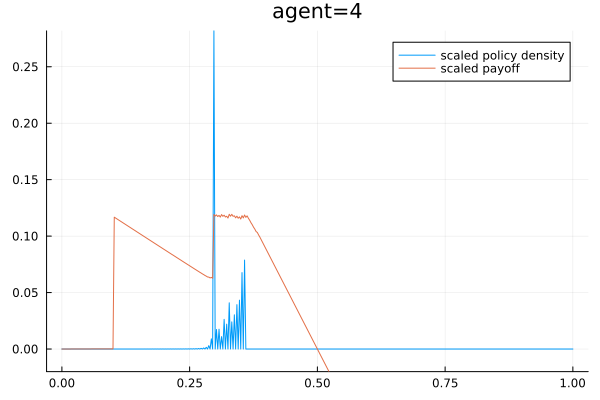}
\includegraphics[width=0.3\textwidth]{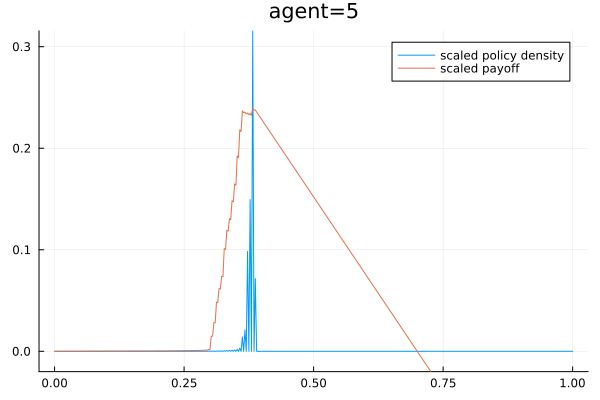}\\
\includegraphics[width=0.3\textwidth]{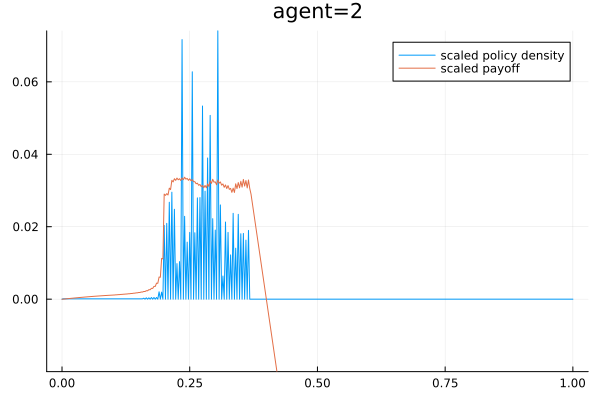}
\includegraphics[width=0.3\textwidth]{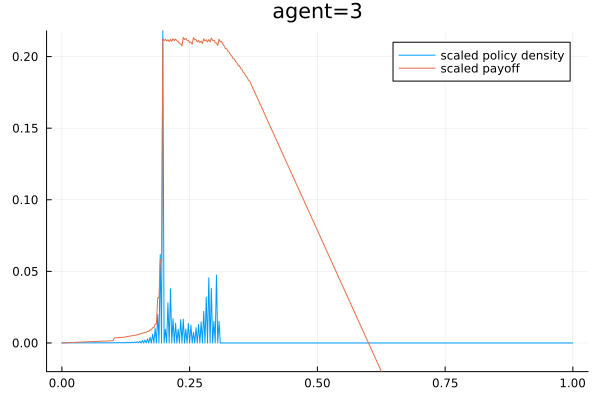}
\includegraphics[width=0.3\textwidth]{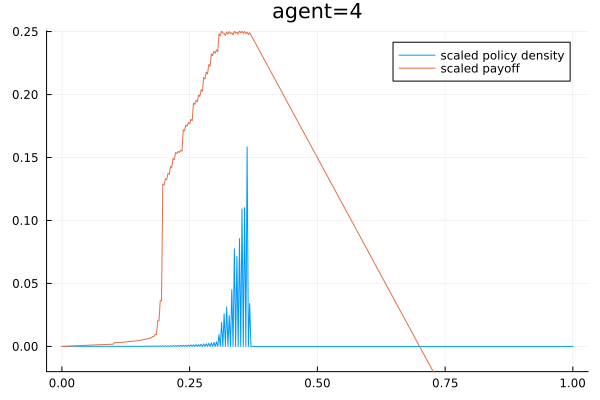}
\\
\includegraphics[width=0.3\textwidth]{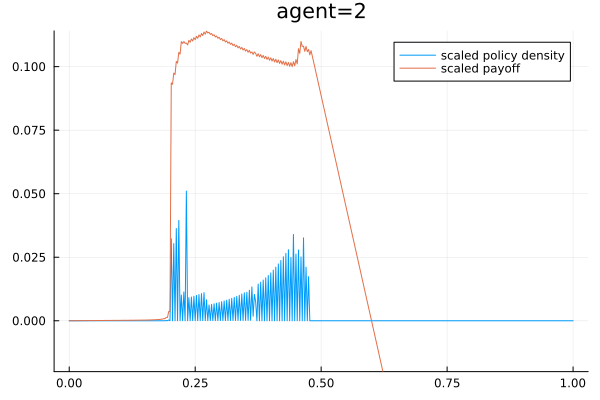}
\includegraphics[width=0.3\textwidth]{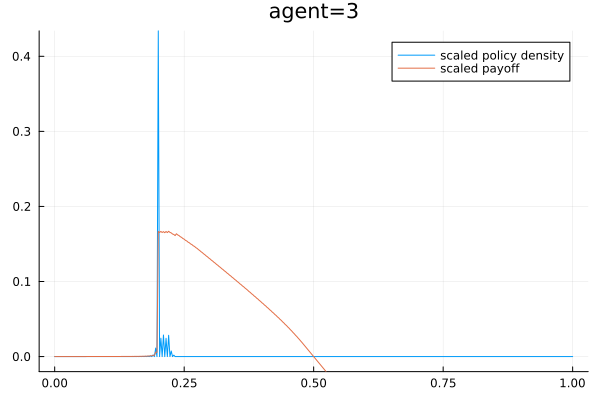}
\includegraphics[width=0.3\textwidth]{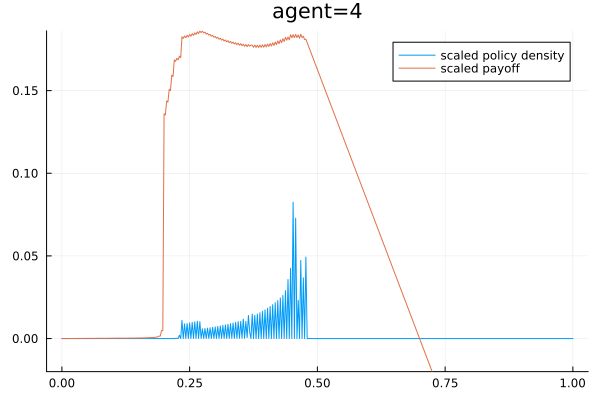}
    \caption{Visual diagnostic of Nash equilibrium convergence. For each agent and auction instance, we superimpose the normalized policy density with the corresponding payoff function. The equilibrium condition—that payoffs are maximal on the policy support—can be verified visually.
    The four rows correspond to different auction configurations:
(Row 1) 4 agents with values $[0.0, 1.0, 0.0, 1.0]$ and scenarios ${[1, 3], [1, 4], [2, 3], [2, 4]}$;
(Row 2) 5 agents with values $[0.2, 0.3, 0.4, 0.5, 0.7]$ and scenarios ${[1, 4], [2, 4], [3, 4], [3, 5], [3, 5]}$;
(Row 3) 4 agents with values $[0.2, 0.4, 0.6, 0.7]$ and scenarios ${[1, 3], [1, 3], [1, 4], [2, 3], [2, 4], [2, 4]}$;
(Row 4) 4 agents with values $[0.2, 0.6, 0.5, 0.7]$ and scenarios ${[1, 3], [1, 3], [1, 4], [2, 3], [2, 4], [2, 4]}$. }
    \label{fig:batch}
\end{figure}

We tested  FP on a symmetric continuous environment, for which there exists a method to compute a numerical solution~\cite{milgrom1982theory}.
The results are shown in Figure~\ref{fig:1}. 
Then, we tested FP on a batch of 20 randomly generated discrete environments with correlated values.
The results are shown in Figure~\ref{fig:2}.
For the two experiments, FP converges to $\epsilon$-equilibria with  a small $\epsilon$. Finally, Figure~\ref{fig:batch} presents results on several auction instances, each characterized by a list of \textsc{values} (one per agent) and a list of \textsc{scenarios}. Since we allow repetition of scenarios, their elements can be treated as equiprobable without loss of generality. To visually assess convergence quality, we overlay each agent's (renormalized) policy density with their corresponding payoffs. At Nash equilibrium, an agent's payoff should be maximal on the support of their policy, making any deviations immediately apparent in these plots.

\begin{thebibliography}{0}
\providecommand{\natexlab}[1]{#1}
\providecommand{\url}[1]{\texttt{#1}}
\expandafter\ifx\csname urlstyle\endcsname\relax
  \providecommand{\doi}[1]{doi: #1}\else
  \providecommand{\doi}{doi: \begingroup \urlstyle{rm}\Url}\fi

\end{thebibliography}


\newpage



\begin{thebibliography}{47}
\providecommand{\natexlab}[1]{#1}
\providecommand{\url}[1]{\texttt{#1}}
\expandafter\ifx\csname urlstyle\endcsname\relax
  \providecommand{\doi}[1]{doi: #1}\else
  \providecommand{\doi}{doi: \begingroup \urlstyle{rm}\Url}\fi

\bibitem[Bajari(2001)]{bajari2001comparing}
Patrick Bajari.
\newblock Comparing competition and collusion: a numerical approach.
\newblock \emph{Economic Theory}, 18\penalty0 (1):\penalty0 187--205, 2001.

\bibitem[Bezanson et~al.(2017)Bezanson, Edelman, Karpinski, and
  Shah]{bezanson2017julia}
Jeff Bezanson, Alan Edelman, Stefan Karpinski, and Viral~B Shah.
\newblock Julia: A fresh approach to numerical computing.
\newblock \emph{SIAM review}, 59\penalty0 (1):\penalty0 65--98, 2017.
\newblock URL \url{https://doi.org/10.1137/141000671}.

\bibitem[Bichler et~al.(2021)Bichler, Fichtl, Heiderkrueger, Kohring, and
  Sutterer]{bichler2021learning}
Martin Bichler, Maximilian Fichtl, S~Heiderkrueger, Nils Kohring, and Paul
  Sutterer.
\newblock Learning equilibria in symmetric auction games using artificial
  neural networks.
\newblock \emph{Nature Machine Intelligence}, 2021.

\bibitem[Brandt et~al.()Brandt, Fischer, and
  Harrenstein]{brandtRateConvergenceFictitious}
Felix Brandt, Felix Fischer, and Paul Harrenstein.
\newblock On the {{Rate}} of {{Convergence}} of {{Fictitious Play}}.

\bibitem[Brown(1951)]{Bro51}
George~W. Brown.
\newblock Iterative solutions of games by fictitious play.
\newblock In T.~C. Coopmans, editor, \emph{Activity Analysis of Productions and
  Allocation}, 374-376. Wiley, 1951.

\bibitem[Cole and Roughgarden(2014)]{cole2014sample}
Richard Cole and Tim Roughgarden.
\newblock The sample complexity of revenue maximization.
\newblock In \emph{Proceedings of the forty-sixth annual ACM symposium on
  Theory of computing}, pages 243--252, 2014.

\bibitem[Dhangwatnotai et~al.(2015)Dhangwatnotai, Roughgarden, and
  Yan]{dhangwatnotai2015revenue}
Peerapong Dhangwatnotai, Tim Roughgarden, and Qiqi Yan.
\newblock Revenue maximization with a single sample.
\newblock \emph{Games and Economic Behavior}, 91:\penalty0 318--333, 2015.

\bibitem[Fibich and Gavish(2011)]{fibich2011numerical}
Gadi Fibich and Nir Gavish.
\newblock Numerical simulations of asymmetric first-price auctions.
\newblock \emph{Games and Economic Behavior}, 73\penalty0 (2):\penalty0
  479--495, 2011.

\bibitem[Fibich and Gavish(2012)]{Fibich2012}
Gadi Fibich and Nir Gavish.
\newblock Asymmetric first-price auctions—a dynamical-systems approach.
\newblock \emph{Mathematics of Operations Research}, 37\penalty0 (2):\penalty0
  219--243, 2012.
\newblock \doi{10.1287/moor.1110.0535}.
\newblock URL \url{https://doi.org/10.1287/moor.1110.0535}.

\bibitem[Filos-Ratsikas et~al.(2021)Filos-Ratsikas, Giannakopoulos, Hollender,
  Lazos, and Po{\c{c}}as]{filos2021complexity}
Aris Filos-Ratsikas, Yiannis Giannakopoulos, Alexandros Hollender, Philip
  Lazos, and Diogo Po{\c{c}}as.
\newblock On the complexity of equilibrium computation in first-price auctions.
\newblock \emph{arXiv preprint arXiv:2103.03238}, 2021.

\bibitem[Fudenberg and Kreps(1993)]{FK93}
Drew Fudenberg and David~M. Kreps.
\newblock Learning mixed equilibria.
\newblock \emph{Games and Economic Behavior}, 5\penalty0 (320-367), 1993.

\bibitem[Ganzfried and Sandholm(2008)]{ganzfried2008computing}
Sam Ganzfried and Tuomas Sandholm.
\newblock Computing an approximate jam/fold equilibrium for 3-player no-limit
  texas hold'em tournaments.
\newblock In \emph{Proceedings of the 7th international joint conference on
  Autonomous agents and multiagent systems-Volume 2}, pages 919--925, 2008.

\bibitem[Gaunersdorfer and Hofbauer(1995)]{GH95}
Andrea Gaunersdorfer and Josef Hofbauer.
\newblock Fictitious play, {Shapley} polygons, and the replicator equation.
\newblock \emph{Games and Economic Behavior}, 11\penalty0 (2):\penalty0
  279--303, 1995.

\bibitem[Gayle and Richard(2008)]{Gayle2008}
Wayne~Roy Gayle and Jean~Francois Richard.
\newblock Numerical solutions of asymmetric, first-price, independent private
  values auctions.
\newblock \emph{Computational Economics}, 32:\penalty0 245--278, 2008.
\newblock ISSN 09277099.
\newblock \doi{10.1007/s10614-008-9125-7}.

\bibitem[Hadikhanloo et~al.(2021)Hadikhanloo, Laraki, Mertikopoulos, and
  Sorin]{HLMS21a}
Saeed Hadikhanloo, Rida Laraki, Panayotis Mertikopoulos, and Sylvain Sorin.
\newblock Learning in nonatomic games, {Part I}: {Finite} action spaces and
  population games.
\newblock \url{https://arxiv.org/abs/2107.01595}, 2021.

\bibitem[Hahn(2008)]{Hah08}
Sunku Hahn.
\newblock The convergence of fictitious play in games with strategic
  complementarities.
\newblock \emph{Economics Letters}, 99\penalty0 (304-306), 2008.

\bibitem[Hartline and Roughgarden(2009)]{hartline2009simple}
Jason~D Hartline and Tim Roughgarden.
\newblock Simple versus optimal mechanisms.
\newblock In \emph{Proceedings of the 10th ACM conference on Electronic
  commerce}, pages 225--234, 2009.

\bibitem[Heymann(2020)]{heymann2020bid}
Benjamin Heymann.
\newblock How to bid in unified second-price auctions when requests are
  duplicated.
\newblock \emph{Operations Research Letters}, 48\penalty0 (4):\penalty0
  446--451, 2020.

\bibitem[Heymann and Mertikopoulos(2021)]{heymann2021heuristic}
Benjamin Heymann and Panayotis Mertikopoulos.
\newblock A heuristic for estimating nash equilibria in first-price auctions
  with correlated values.
\newblock \emph{arXiv preprint arXiv:2108.04506}, 2021.

\bibitem[Hofbauer(1995)]{Hof95b}
Josef Hofbauer.
\newblock Stability for the best response dynamics.
\newblock mimeo, 1995.

\bibitem[Hofbauer and Sandholm(2009)]{HS09}
Josef Hofbauer and William~H. Sandholm.
\newblock Stable games and their dynamics.
\newblock \emph{Journal of Economic Theory}, 144\penalty0 (4):\penalty0
  1665--1693, July 2009.

\bibitem[Hubbard and Paarsch(2014)]{hubbard2014numerical}
Timothy~P Hubbard and Harry~J Paarsch.
\newblock On the numerical solution of equilibria in auction models with
  asymmetries within the private-values paradigm.
\newblock In \emph{Handbook of computational economics}, volume~3, pages
  37--115. Elsevier, 2014.

\bibitem[Jordan(1993)]{Jor93}
James~S. Jordan.
\newblock Three problems in learning mixed strategy {Nash} equilibria.
\newblock \emph{Games and Economic Behavior}, 5\penalty0 (3):\penalty0
  368--386, July 1993.

\bibitem[Kaplan and Zamir(2012)]{kaplan2012asymmetric}
Todd~R Kaplan and Shmuel Zamir.
\newblock Asymmetric first-price auctions with uniform distributions: analytic
  solutions to the general case.
\newblock \emph{Economic Theory}, 50\penalty0 (2):\penalty0 269--302, 2012.

\bibitem[Krishna(1992)]{Kri92}
Vijay Krishna.
\newblock Learning in games with strategic complementarities.
\newblock mimeo, 1992.

\bibitem[Krishna(2009)]{krishna2009auction}
Vijay Krishna.
\newblock \emph{Auction theory}.
\newblock Academic press, 2009.

\bibitem[Lebrun(1999)]{10.2307/2648842}
Bernard Lebrun.
\newblock First price auctions in the asymmetric n bidder case.
\newblock \emph{International Economic Review}, 40\penalty0 (1):\penalty0
  125--142, 1999.
\newblock ISSN 00206598, 14682354.
\newblock URL \url{http://www.jstor.org/stable/2648842}.

\bibitem[Marshall et~al.(1994)Marshall, Meurer, Richard, and
  Stromquist]{marshall1994numerical}
Robert~C Marshall, Michael~J Meurer, Jean-Francois Richard, and Walter
  Stromquist.
\newblock Numerical analysis of asymmetric first price auctions.
\newblock \emph{Games and Economic Behavior}, 7\penalty0 (2):\penalty0
  193--220, 1994.

\bibitem[Maskin and Riley(2003)]{maskin2003uniqueness}
Eric Maskin and John Riley.
\newblock Uniqueness of equilibrium in sealed high-bid auctions.
\newblock \emph{Games and Economic Behavior}, 45\penalty0 (2):\penalty0
  395--409, 2003.

\bibitem[Mertikopoulos and Sandholm(2016)]{MS16}
Panayotis Mertikopoulos and William~H. Sandholm.
\newblock Learning in games via reinforcement and regularization.
\newblock \emph{Mathematics of Operations Research}, 41\penalty0 (4):\penalty0
  1297--1324, November 2016.

\bibitem[Milgrom and Roberts(1990)]{MR90}
Paul Milgrom and John Roberts.
\newblock Rationalizability, learning, and equilibrium in games with strategic
  complementarities.
\newblock \emph{Econometrica}, 58\penalty0 (6):\penalty0 1255--1277, November
  1990.

\bibitem[Milgrom and Roberts(1991)]{MR91}
Paul Milgrom and John Roberts.
\newblock Adaptive and sophisticated learning in normal form games.
\newblock \emph{Games and Economic Behavior}, 3:\penalty0 82--100, 1991.

\bibitem[Milgrom and Weber(1982)]{milgrom1982theory}
Paul~R Milgrom and Robert~J Weber.
\newblock A theory of auctions and competitive bidding.
\newblock \emph{Econometrica: Journal of the Econometric Society}, pages
  1089--1122, 1982.

\bibitem[Miyasawa(1961)]{Miy61}
K.~Miyasawa.
\newblock On the convergence of learning processes in a $2 \times 2$ non-zero
  sum game.
\newblock Research memorandum~33, Princeton University, 1961.

\bibitem[Monderer and Shapley(1996)]{MS96-jet}
Dov Monderer and Lloyd~S. Shapley.
\newblock Fictitious play property for games with identical interests.
\newblock \emph{Journal of Economic Theory}, 68:\penalty0 256--265, 1996.

\bibitem[Myerson(1981)]{myerson1981optimal}
Roger~B Myerson.
\newblock Optimal auction design.
\newblock \emph{Mathematics of operations research}, 6\penalty0 (1):\penalty0
  58--73, 1981.

\bibitem[Paes~Leme et~al.(2020)Paes~Leme, Sivan, and Teng]{paes2020competitive}
Renato Paes~Leme, Balasubramanian Sivan, and Yifeng Teng.
\newblock Why do competitive markets converge to first-price auctions?
\newblock In \emph{Proceedings of The Web Conference 2020}, pages 596--605,
  2020.

\bibitem[Plum(1992)]{plum1992characterization}
Michael Plum.
\newblock Characterization and computation of nash-equilibria for auctions with
  incomplete information.
\newblock \emph{International Journal of Game Theory}, 20\penalty0
  (4):\penalty0 393--418, 1992.

\bibitem[Rabinovich et~al.(2013)Rabinovich, Naroditskiy, Gerding, and
  Jennings]{rabinovich2013computing}
Zinovi Rabinovich, Victor Naroditskiy, Enrico~H Gerding, and Nicholas~R
  Jennings.
\newblock Computing pure bayesian-nash equilibria in games with finite actions
  and continuous types.
\newblock \emph{Artificial Intelligence}, 195:\penalty0 106--139, 2013.

\bibitem[Rasooly and Gavidia-Calderon(2021)]{rasooly2021importance}
Itzhak Rasooly and Carlos Gavidia-Calderon.
\newblock The importance of being discrete: on the inaccuracy of continuous
  approximations in auction theory, 2021.

\bibitem[Reny and Zamir(2004)]{reny2004existence}
Philip~J Reny and Shmuel Zamir.
\newblock On the existence of pure strategy monotone equilibria in asymmetric
  first-price auctions.
\newblock \emph{Econometrica}, 72\penalty0 (4):\penalty0 1105--1125, 2004.

\bibitem[Robinson(1951)]{Rob51}
Julia Robinson.
\newblock An iterative method for solving a game.
\newblock \emph{Annals of Mathematics}, 54:\penalty0 296--301, 1951.

\bibitem[Roughgarden(2016)]{Roughgarden_2016}
Tim Roughgarden.
\newblock \emph{Twenty Lectures on Algorithmic Game Theory}.
\newblock Cambridge University Press, 2016.

\bibitem[Shalev-Shwartz(2011)]{SS11}
Shai Shalev-Shwartz.
\newblock Online learning and online convex optimization.
\newblock \emph{Foundations and Trends in Machine Learning}, 4\penalty0
  (2):\penalty0 107--194, 2011.

\bibitem[Shapley(1964)]{Sha64}
Lloyd~S. Shapley.
\newblock Some topics in two-person games.
\newblock In \emph{Advances in Game Theory}, number~52 in Annals of Mathematics
  Studies. Princeton University Press, 1964.

\bibitem[Vickrey(1961)]{vickrey1961counterspeculation}
William Vickrey.
\newblock Counterspeculation, auctions, and competitive sealed tenders.
\newblock \emph{The Journal of finance}, 16\penalty0 (1):\penalty0 8--37, 1961.

\bibitem[Wang et~al.(2020)Wang, Shen, and Zuo]{10.5555/3398761.3398929}
Zihe Wang, Weiran Shen, and Song Zuo.
\newblock Bayesian nash equilibrium in first-price auction with discrete value
  distributions.
\newblock In \emph{Proceedings of the 19th International Conference on
  Autonomous Agents and MultiAgent Systems}, AAMAS '20, page 1458–1466,
  Richland, SC, 2020. International Foundation for Autonomous Agents and
  Multiagent Systems.
\newblock ISBN 9781450375184.

\end{thebibliography}
\end{document}